\documentclass[12pt]{iopart}
\usepackage{graphicx}
\usepackage{amssymb}

\newcommand{\beq}{\begin{eqnarray}}
\newcommand{\eeq}{\end{eqnarray}}

\begin{document}

\title{Solving the Helmholtz equation for membranes of arbitrary shape}
\author{Paolo Amore\footnote{paolo.amore@gmail.google.com}}
\address{Facultad de Ciencias, Universidad de Colima, \\
Bernal D\'{\i}az del Castillo 340, Colima, Colima, M\'exico} 
\address{and Physics Department, University of Texas at El Paso, \\
El Paso, Texas, USA}

\begin{abstract}
I calculate the modes of vibration of membranes of arbitrary shape using a collocation 
approach based on Little Sinc Functions. The matrix representation of the PDE obtained
using this method is explicit and it does not require the calculation of integrals.
To illustrate the virtues of this approach, I have considered a large number of examples,
part of them taken from the literature, and part of them new. When possible, I have tested
the accuracy of these results by comparing them with the exact results (when available) or 
with results from the literature. In particular, in the case of the L-shaped membrane, 
the first example discussed in the paper, I show that it is possible to extrapolate the 
results obtained with different grid sizes to obtain higly precise results. Finally, I also 
show that the present collocation technique can be easily combined with conformal mapping 
to provide numerical approximations to the energies which quite rapidly converge to the exact
results.
\end{abstract}

\maketitle

\section{Introduction}
\label{intro}

This paper considers the problem of solving the Helmholtz equation 
\beq
- \Delta \psi(x,y) = E \psi(x,y)
\eeq
over a two-dimensional domain, $B$, of arbitrary shape, assuming Dirichlet boundary conditions over
the border, $\partial B$. Physically, this equation describes the classical vibration of a homogenoeous membrane
or the behaviour of a particle confined in a region with infinite walls in quantum mechanics.
Unfortunately exact solutions to this equation are available only in few cases, such as for a rectangular or a 
circular membrane, where they can be expressed in terms of trigonometric and Bessel functions 
respectively~\cite{FW80}. In the majority of cases, in fact, only numerical approaches can be used: some of these 
approaches are discussed for example in a beautiful paper by Kuttler and Sigillito, \cite{KS84}. 
The purpose of the present paper is to introduce a different approach to the numerical solution of the Helmholtz equation 
(both homogenous and inhomogeneous) and illustrate its strength and flexibility by applying it to a large number 
of examples. 

The paper is organized as follows: in Section \ref{method} I describe the method and discuss its application to the
classical problem of a L-shaped membrane; in Section \ref{africa} I consider an homogenous membrane, with the shape of Africa and
calculate few states; in Section \ref{isospec} I consider two inequivalent membranes, which are known to be isospectral, 
obtaining a numerical indication of isospectrality; in Section \ref{unusualdrum} I study an example of irregular drum;
in Section \ref{BIC} the method is applied to study the emergence of bound states in a configuration of wires of 
neglegible transverse dimension, in presence of crossings; in Section \ref{conformal} I show that even more precise results
can be achieved by combining the collocation method with a conformal mapping of the boundary. Finally, in Section \ref{conclusion}
I draw my conclusions.

\section{The method}
\label{method}

The method that I propose in this paper uses a particular set of functions,
the {\sl Little Sinc functions} (LSF) of \cite{Amore07a,Amore07b}, to obtain a
discretization of a finite region of the two-dimensional plane. These functions
have been used with success in the numerical solution of the Schr\"odinger equation
in one dimension, both for problems restricted to finite intervals and for problems
on the real line. In particular it has been proved that exponential convergence
to the exact solution can be reached when variational considerations are made 
(see \cite{Amore07a,Amore07b}).

Although Ref.\cite{Amore07a} contains a detailed discussion of the LSF, I will briefly
review here the main properties, which will be useful in the paper. Throughout the paper I will
follow the notation of \cite{Amore07a}.

A Little Sinc Function is obtained as an approximate representation of the Dirac delta function
in terms of the wave functions of a particle in a box (being $2L$ the size of the box). Straightforward
algebra leads to the expression
\beq
s_k(h,N,x) \equiv  \frac{1}{2 N} \ \left\{ \frac{\sin \left( (2N+1) \ \chi_-(x)\right)}{\sin \chi_-(x)}
-\frac{\cos\left((2N+1)  \chi_+(x)\right)}{\cos \chi_+(x)} \right\} \ .
\label{sincls}
\eeq
where $\chi_\pm(x) \equiv \frac{\pi}{2 N h} (x \pm k h)$. The index $k$ takes the integer values
between $-N/2+1$ and $N/2-1$ ($N$ being an even integer). The LSF corresponding to a specific 
value of $k$ is peaked at $x_k = 2 L k/N = k h$, $h$ being the grid spacing and $2L$ the total extension
of the interval where the function is defined. By direct inspection of eq.~(\ref{sincls}) it is found that
$s_k(h,N,x_j) = \delta_{kj}$, showing that the LSF takes its maximum value at the $k^{th}$ grid 
point and vanishes on the remaining points of the grid.

It can be easily proved that the different LSF corresponding to the same set are orthogonal~\cite{Amore07a}:
\beq
\int_{-L}^L s_k(h,N,x) s_j(h,N,x) dx = h \ \delta_{kj} 
\eeq
and that a function defined on $x\in (-L,L)$ may be approximated as 
\beq
f(x) \approx \sum_{k=-N/2+1}^{N/2-1} f(x_k) \ s_k(h, N,x) \ . \label{f(x)_LSF}
\eeq

This formula can be applied to obtain a representation of the derivative of a LSF in terms of the set of LSF
as:
\beq
\frac{d s_k(h,N,x)}{dx} \approx \sum_j \left. \frac{d s_k(h,N,x)}{dx}\right|_{x=x_j} \ s_j(h,N,x) 
\equiv \sum_j c_{kj}^{(1)} \ s_j(h,N,x) \nonumber \\
\frac{d^2 s_k(h,N,x)}{dx^2} \approx \sum_j  \left. \frac{d^2 s_k(h,N,x)}{dx^2}\right|_{x=x_j} \ s_j(h,N,x) \equiv 
\sum_j c_{kj}^{(2)} \ s_j(h,N,x)  \ , \nonumber \\
\label{der2}
\eeq
where the expressions for the coefficients $c_{kj}^{(r)}$ can be found in \cite{Amore07a}. Although eqs.(\ref{f(x)_LSF})
is approximate and the LSF strictly speaking do not form a basis, the error made with this approximation decreases with $N$ and tends to
zero as $N$ tends to infinity, as shown in \cite{Amore07a}. For this reason, the effect of this approximation is 
essentially to replace the continuum of a interval of size $2L$ on the real line with a discrete set of $N-1$ points, $x_k$,
uniformly spaced on this interval. 

Clearly these relations are easily generalized to functions of two or more variables. Since the focus of this paper is
on two dimensional membranes, I will briefly discuss how the LSF are used to discretize a region of the plane; the
extension to higher dimensional spaces is straightforward.  
A function of two variables can be approximated in terms of $(N_x-1) \times (N_y-1)$ functions, corresponding to  
the direct product of the $N_x-1$ and $N_y-1$ LSF in the $x$ and $y$ axis: each term in this set corresponds to a specific 
point on a rectangular grid with spacings $h_x$ and $h_y$ (in this paper I use a square grid with $N_x=N_y=N$ and $L_x = L_y=L$).

Since $(k,k')$ identifies a unique point on the grid, I can select this point using a single index
\beq
K &\equiv& k'+\frac{N}{2} + (N-1) \left(k+\frac{N}{2}-1\right) \label{KK} 
\eeq
which can take the values $1 \leq K \leq (N-1)^2$. I can also invert this relation and write
\beq
k &=& 1 - N/2 + \left[\frac{K}{N-1+\varepsilon}\right] \\ 
k' &=& K - N/2 - (N-1) \ \left[ \frac{K}{N-1+\varepsilon}\right] \ ,
\eeq
where $\left[ a \right]$ is the integer part of a real number $a$ and $\varepsilon \rightarrow 0$.

As a natural extension of the results presented in \cite{Amore07a,Amore07b} I can consider the 
Schr\"odinger equation in two dimensions
\beq
\hat{H} \psi_n(x,y) \equiv \left[- \Delta + V(x,y) \right] \psi_n(x,y) = E_n \psi_n(x,y) 
\label{schrodinger}
\eeq
using the convention of assuming a particle of mass $m=1/2$ and setting $\hbar = 1$. The Helmholtz
equation, which describes the vibration of a membrane, is a special case of (\ref{schrodinger}), corresponding
to having $V(x,y) = 0$ inside the region ${\cal B}$ where the membrane lies and $V(x,y)=\infty$ on the border $\partial {\cal B}$
and outside the membrane. 

The discretization of eq.~(\ref{schrodinger}) proceeds in a simple way using the properties 
discussed in eqs.~(\ref{f(x)_LSF}) and (\ref{der2}):
\beq
H_{kk',jj'} =  -\left[ c^{(2)}_{kj} \delta_{k'j'} + \delta_{kj} c^{(2)}_{k'j'} \right]+ \delta_{kj} \delta_{k'j'} V(x_k,y_{k'}) 
\label{Hamiltonian}
\eeq
where $(k,j,k',j') = - N/2+1, \dots, N/2-1$. Notice that the potential part of the Hamiltonian is obtained 
by simply "collocating" the potential $V(x,y)$ on the grid, an operation with a limited computational
price. The result shown in (\ref{Hamiltonian}) corresponds to the matrix element of the Hamiltonian operator
$\hat{H}$ between two grid points, $(k,k')$ and $(j,j')$, which can be selected using two integer values $K$ and $J$, 
as shown in (\ref{KK}). 

Following this procedure the solution of the Schr\"odinger (Helmholtz) equation on the uniform grid generated by the LSF
corresponds to the diagonalization of a $(N-1)^2 \times (N-1)^2$ square matrix, whose elements are given by eq.~(\ref{Hamiltonian}).

\begin{figure}[t]
\begin{center}
\includegraphics[width=6cm]{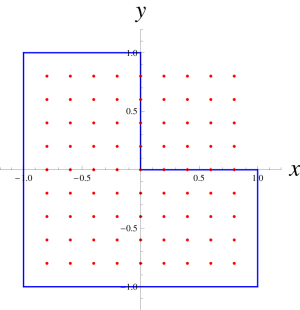}
\caption{L shaped membrane. The dots are the collocation
points corresponding to $N=14$.}
\label{fig_lshape}
\end{center}
\end{figure}

I will now use a specific problem, the vibration of a L-shaped membrane, represented in Fig.\ref{fig_lshape}, 
to illustrate the method, and discuss different implementations of the method itself.  
This problem has been widely used in the past to test the performance of the different numerical methods (see for example
refs. \cite{RW65,FHM67,Mas67,KS84,Sid84,Schiff88,PlDr04,BT05,TB06}) and it is therefore a useful tool to assess the strength 
of the present approach. Because of the reentrant corner, corresponding to the angle $\theta =3 \pi/2$ located at $(0,0)$, the derivatives
of $\psi(x,y)$ in the radial direction are unbounded (see \cite{RW65}).  

Reid and Walsh in \cite{RW65} obtained a numerical approximation for the two lowest modes of this membrane using  finite 
differences  and a confomal map which eliminates the reentrant corner (see fig.5 of \cite{RW65});
a more precise result was later obtained by Fox, Henrici and Moler who used the Method of Particular Solutions (MPS) in \cite{FHM67} exploiting
the symmetries of the problem (the reader may find a detailed  discussion of the symmetries for this problem in \cite{KS84}):  
the first eight digits of the lowest eigenvalue reported by the authors are correct. 
Mason has obtained numerical estimates for the first few modes of the L-shaped membrane in terms of a two dimensional Chebyshev series \cite{Mas67}.
Milsted and Hutchinson \cite{MH73} have obtained finite element solutions to this problem.
Sideridis in \cite{Sid84} used a conformal mapping of the L-shaped region onto a square and then solved the resulting equation on a
uniform rectangular mesh, obtaining the first four digits of the lowest mode. Schiff, ref. \cite{Schiff88}, has calculated
the first $15$ lowest modes of this membrane using finite elements, with a refined grid covering the region surrounding the reentrant corner.

More recently Platte and Driscoll have solved the boundary value problem on the L-shape membrane using radial basis functions ~\cite{PlDr04}.
Finally Betcke and Trefethen have revisited the MPS in \cite{BT05}; in that paper they have observed that the MPS reaches a minimal 
error for a certain value of $N$ (the number of collocation points on each of the sides non adjacent 
to the corner where the expansion is performed) but then it starts to grow as $N$ increases. 
The modified version of the method discussed in \cite{BT05}, which samples the Fourier-Bessel 
functions also in the interior points, corrects this problem and provides a convergent behaviour 
for the error. In this way Betcke and Trefethen were able to obtain the first $14$ digits of 
the lowest eigenvalue of the L-shaped membrane, $E_1 \approx 9.6397238440219$. I will use 
this precise result to test the accuracy of our method. Ref.\cite{TB06} contains precise estimates for some higher excited states
of the L-shaped membrane.

\begin{table}
\begin{center}
\begin{tabular}{|ccccccccc|}
\hline
$n$ &  $E_n^{(-)}$ & $E_n^{(+)}$ & $n$ &  $E_n^{(-)}$ & $E_n^{(+)}$ & $n$ &  $E_n^{(-)}$ & $E_n^{(+)}$ \\
\hline
1  &   9.177164983  &   9.725740015  &   37  &   178.4844965  &   184.8421046  &   73  &   343.7275747  &     361.2225873\\
2  &   14.78073926  &   15.25792488  &   38  &   193.8672234  &   197.9262249  &   74  &   344.9934435  &     361.9427155\\
3  &   19.37069304  &   19.79218759  &   39  &   193.8814147  &   197.9267047  &   75  &   358.2837562  &     366.3264978\\
4  &   29.22316338  &   29.5638567  &   40  &   198.1817051  &   201.9430968  &   76  &   359.5348189  &     366.3670805\\
5  &   30.96354609  &   32.09126661  &   41  &   201.4624222  &   208.1352708  &   77  &   364.75956  &     370.6707929\\
6  &   40.02046425  &   41.71342235  &   42  &   208.0809373  &   208.8812769  &   78  &   365.5537093  &     373.5349023\\
7  &   43.35567534  &   45.16667725  &   43  &   208.3091478  &   209.9202486  &   79  &   367.4408345  &     377.7859531\\
8  &   48.49170563  &   49.48205954  &   44  &   218.9529607  &   223.795143  &   80  &   371.1572423  &     384.5448247\\
9  &   48.50129154  &   49.48210584  &   45  &   219.7977882  &   224.2020086  &   81  &   380.3860374  &     390.8564172\\
10  &   55.00253452  &   56.99285853  &   46  &   230.4881589  &   237.0180886  &   82  &   381.619838  &     391.5359069\\
11  &   64.39311656  &   65.51743185  &   47  &   234.522406  &   238.9225242  &   83  &   389.8803312  &     395.9403897\\
12  &   70.17580289  &   71.22539692  &   48  &   240.8305192  &   247.399449  &   84  &   390.1339997  &     395.9528626\\
13  &   70.75536576  &   71.694315  &   49  &   241.3085263  &   247.4007186  &   85  &   396.7580769  &     405.7350508\\
14  &   77.43821507  &   79.16827278  &   50  &   242.091938  &   251.9722965  &   86  &   396.783386  &     405.7417585\\
15  &   85.62358216  &   89.95280767  &   51  &   242.4936924  &   253.9585512  &   87  &   405.4299759  &     417.4274921\\
16  &   89.20133569  &   92.66479784  &   52  &   252.6896353  &   257.3835642  &   88  &   409.020823  &     421.0196083\\
17  &   95.02656477  &   97.67618899  &   53  &   254.3713602  &   257.4891189  &   89  &   422.9007962  &     426.9812301\\
18  &   96.44902117  &   98.9581845  &   54  &   258.45965  &   267.2165568  &   90  &   425.5099017  &     427.4264825\\
19  &   97.50939511  &   98.98841467  &   55  &   262.2523481  &   270.362378  &   91  &   425.6809979  &     434.295792\\
20  &   99.15370293  &   102.1148968  &   56  &   262.9019839  &   271.2576246  &   92  &   435.8786896  &     444.6206086\\
21  &   109.8094028  &   112.7440906  &   57  &   276.4727322  &   282.0671402  &   93  &   435.9192729  &     445.3287181\\
22  &   112.6706295  &   115.940658  &   58  &   279.3270569  &   286.9614486  &   94  &   438.0162243  &     445.3290091\\
23  &   125.8637839  &   128.647868  &   59  &   281.7073779  &   287.0144046  &   95  &   442.0953758  &     454.6858173\\
24  &   126.0084139  &   128.6517412  &   60  &   284.1271564  &   290.2236089  &   96  &   449.9029921  &     454.7270884\\
25  &   129.4077703  &   130.2193886  &   61  &   287.835501  &   294.4411423  &   97  &   450.2786574  &     456.2838516\\
26  &   129.4610293  &   130.4087363  &   62  &   301.6031656  &   306.989597  &   98  &   455.431713  &     465.7654543\\
27  &   138.4732345  &   143.0937626  &   63  &   304.0018517  &   307.2227265  &   99  &   463.9179935  &     480.155258\\
28  &   148.8908462  &   151.4047394  &   64  &   307.1623397  &   311.2775619  &   100  &   464.4999312    &   481.0156905\\
29  &   149.3132131  &   155.4149531  &   65  &   308.637254  &   314.7477734  &   101  &   468.8725774  &     488.2586507\\
30  &   157.1294641  &   162.6916706  &   66  &   310.2539585  &   316.6847591  &   102  &   471.7598792    &   491.8389757\\
31  &   159.2280728  &   165.3935921  &   67  &   328.509336  &   336.1784596  &   103  &   478.7708593  &     494.7920322\\
32  &   159.647457  &   165.4142765  &   68  &   328.7797637  &   336.2277068  &   104  &   484.8647328  &     494.949551\\
33  &   166.1106112  &   168.2997856  &   69  &   330.9763129  &   336.5300454  &   105  &   485.0320322    &   495.6089426\\
34  &   166.1907708  &   168.3155339  &   70  &   332.6951225  &   336.5831701  &   106  &   493.2198808    &   501.7272125\\
35  &   173.3430437  &   178.1220953  &   71  &   340.1437751  &   346.949885  &   107  &   493.4054114  &     503.8197404\\
36  &   175.4511857  &   180.5827508  &   72  &   340.2548393  &   353.4120482  &   108  &   499.6245389    &   514.1288667\\
\hline
\end{tabular}
\label{tableL}
\end{center}
\bigskip
\caption{First 108 eigenvalues of the L-shaped membrane calculated with a grid with $N=60$. }
\end{table}

I will now apply the LSF to the numerical solution of this problem: looking at Fig.~\ref{fig_lshape} 
I consider the grid points which are internal to the membrane and which do not fall on the border.
For a fixed $N$ there is a total of  $3/4 N^2-2 N+1$ points; the grid represented in the figure 
corresponds to $N=14$ and therefore to a total of $120$ internal points. 
In this case the collocation of the Hamiltonian on the uniform grid generated by the LSF leads 
to a $120 \times 120$ matrix, which can then be diagonalized. The eigenvalues of this matrix provide 
the lowest $120$ modes of the membrane, while the eigenvectors provide the lowest $120$ wave functions.
Alternatively I can pick all the points of the grid internal to the membrane, including those falling on the border:
in such case a total of $3/4 N^2-N$ points is found, corresponding to a total of
$133$ points in the case of the Figure.

Table \ref{tableL} contains the first $108$ eigenvalues of the L-shaped membrane calculated using a grid with $N=60$ 
and selecting the grid points according to the prescriptions just explained. I have used the notation $E_n^{(\pm)}$ for
the energy of the $n^{th}$ state when the collocation points on the border are either rejected ($E_n^{(+)}$) or kept 
($E_n^{(-)}$). The notation $(\pm)$ is used since the two sets approach the exact results either from above ($+$) or 
from below ($-$), as one can see comparing these numbers with the precise results contained in \cite{BT05,TB06}. 
The reader will certainly notice that the results of Table \ref{tableL} contain rather large errors: in the case of the 
fundamental state, for example, one has an error of about $1 \%$ from $E_n^{(+)}$ and a much larger error of almost $5 \%$ 
for $E_n^{(-)}$. 

\begin{figure}[ht]
\begin{center}
\includegraphics[width=7cm]{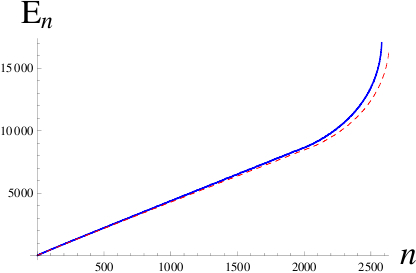}
\hspace{.5cm}
\includegraphics[width=7cm]{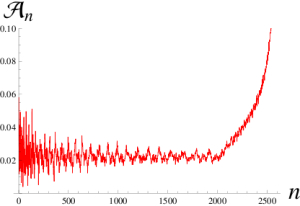}
\end{center}
\caption{Left panel:Energy of the ground state of the L shaped membrane as a function of the number of grid points $N$. 
The horizontal line is the precise result of \cite{BT05}. The set approaching the exact result from above (below) 
corresponds to $E_1^{(+)}$ ($E_1^{(-)}$). Right panel: The asymmetry ${\cal A}_n = 2 (E_n^{(+)}-E_n^{(-)})/(E_n^{(+)}+E_n^{(-)})$ 
calculated with a grid with $N=60$.}
\label{fig_updown}
\end{figure}

The left panel of Fig.\ref{fig_updown} shows the eigenvalues $E_n^{(+)}$ (solid line) and $E_n^{(-)}$ (dashed line) 
for the L-shaped membrane corresponding to a grid with $N=60$.
The reader may notice that the higher end of the spectrum displays a curvature, contrary to the behaviour
predicted by Weyl's law, i.e. $\langle N \rangle \propto E$ for large energies. It is easy to show 
that such effect is artificial: consider for example the case of a particle confined in a unit square, whose 
energies are given by $E_{n_x,n_y}= (n_x^2+n_y^2) \pi^2$. The diagonalization of the Hamiltonian (\ref{Hamiltonian})
for this problem would provide the energies corresponding to the $(N-1)^2$ states obtained taking the first 
$N-1$ values of $n_x$ and $n_y$. This means that for energies higher than $E_{N} = \left[N^2-2 N+2\right] \pi^2$ 
the method will provide only the eigenvalues contained inside a square of side $N-1$ (in the $(n_x,n_y)$ plane),
up to a maximal energy $E_{MAX} = 2 \left[N^2-2 N+2\right] \pi^2$. For this reason, the states above $E_N$
are incomplete and should not be taken into account for inferring the asymptotic behavior of $\langle N\rangle$.
The right panel of Fig.\ref{fig_updown} displays the asymmetry defined as ${\cal A}_n = 2 (E_n^{(+)}-E_n^{(-)})/(E_n^{(+)}+E_n^{(-)})$
for the same grid: this quantity provides an upper estimate for the error.

\begin{figure}[t]
\begin{center}
\includegraphics[width=8cm]{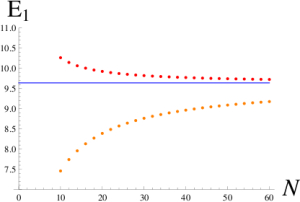}
\caption{Energy of the ground state of the L shaped membrane as a function of the number of grid points $N$. 
The horizontal line is the precise result of \cite{BT05}. The set approaching the exact result from above (below) 
corresponds to $E_1^{(+)}$ ($E_1^{(-)}$).}
\label{fig_energy}
\end{center}
\end{figure}

Fig. \ref{fig_energy} displays the ground state energy of the L-shaped membrane as a function 
of the number of grid points and compares it with the precise result of \cite{BT05}: as already pointed out the two sets
approach the exact value from above and below.

Much more precise results can be obtained by performing an extrapolation of the results corresponding
to finite grids: this is a common procedure used in the literature (see for example \cite{KS84}). 
I have considered four different extrapolation sets using the numerical results obtained working with grids with $N$ ranging
from $N=10$ to $N=60$ (only even values). Calling $h=2L/N$ the grid spacing the sets are:
\beq
f_1(h) = \sum_{n=0}^{\bar{N}} c_n h^n \label{fit1} \\
f_{2}(h) = \frac{\sum_{n=0}^{\bar{N}/2} c_n h^n}{1+\sum_{n=1}^{\bar{N}/2} c_n h^n} \label{fit2} \\
f_{3}(h) = c_0+\sum_{n=1}^{\bar{N}} c_n h^{n/3+2/3} \label{fit3} \\
f_{4}(h) = \frac{c_0+\sum_{n=1}^{\bar{N}/2} c_n h^{n/3+2/3}}{1+\sum_{n=1}^{\bar{N}/2} c_n h^{n/3+2/3}} \label{fit4}
\eeq
where $\bar{N}$ is an even integer which determines the number of coefficients used in the fits.

The continuum limit is reached taking $h \rightarrow \infty$, where only the coefficient
$c_0$ survives. The unknown coefficients in the expressions (\ref{fit1}), (\ref{fit2}), 
(\ref{fit3}) and (\ref{fit4}) are obtained using a Least Square approach: I show the 
results of this procedure in Table \ref{tableL2}. In general, the last set provides 
the best results and indeed it reproduces the first $11$ digits of $E_1$ correctly, 
using either the values of $E_1^{(-)}$ or those of $E_1^{(+)}$. In the case of $E_3$, for which 
the exact value is known ($E_3=2 \pi^2$), I obtain the first $14$ digits correct using $E_3^{(+)}$
and the first $11$ digits correct using $E_3^{(-)}$. 

\begin{table}
\begin{center}
\begin{tabular}{|c|c|cccc|}
\hline
 $n$ & & Set 1 & Set 2 & Set 3 & Set 4 \\
\hline
1  & ${(-)}$ & 9.63959383529194 & 9.63970774930113 & 9.63972385784876 & 9.63972384404696$^*$ \\
1  & ${(+)}$ & 9.63959513453456 & 9.63971258279395 & 9.63972384034031 & 9.63972384401891$^*$ \\
2  & ${(-)}$ & 15.1972518419212 & 15.1974702475024 & 15.1972519362081 & 15.1972519266011$^*$ \\
2  & ${(+)}$ & 15.1972518428845 & 15.1972519235114 & 15.1972519387503 & 15.1972519264561$^*$ \\
3  & ${(-)}$ & 19.7392087861784 & 19.7392088017282 & 19.7392073765870 & 19.7392088020095$^*$ \\
3  & ${(+)}$ & 19.7392088019879 & 19.7392088021704 & 19.7392087962239 & 19.7392088021785$^*$ \\
4  & ${(-)}$ & 29.5178267971821 & 29.5214811097206 & $-$ & 29.5214811103487$^*$ \\
4  & ${(+)}$ & 29.5214810813053 & 29.5214811126514 & 29.5214794563921 & 29.5214811141506$^*$ \\
5  & ${(-)}$ & 31.9159767579531 & 31.9125745966885 & $-$ & 31.9126359533035$^*$\\
5  & ${(+)}$ & 31.9123209946513 & 31.9126005580344 & 31.9126386707453 & 31.9126359571263$^*$ \\
6  & ${(-)}$ & 41.474267306813  & 41.4744740922213 & 41.4761914432832 & 41.4745099148779$^*$ \\
6  & ${(+)}$ & 41.4742739974452 & 41.4744780007070 & 41.4741677038785 & 41.4745098904487$^*$ \\
20 & ${(-)}$ & 101.776561675314$^*$ & 101.605333389975 & - & 99.7713224851033 \\
20 & ${(+)}$ & 101.604853531780 & 101.605223692426 & 101.673183488214 & 101.605294080845$^*$ \\
50 & ${(-)}$ & - & 246.740564791939 & - & 246.602432808866$^*$ \\
50 & ${(+)}$ & 250.784799377301 & 250.785244396338 & - & 250.785494606618$^*$ \\
104 & ${(-)}$ & - & 410.08260648211 & - & - \\
104 & ${(+)}$ & 493.480067984180$^*$ & 493.480206216096 &  -& 493.488405725447 \\
\hline
\end{tabular}
\label{tableL2}
\end{center}
\bigskip
\caption{Extrapolation of the nine eigenvalues of the L-shaped membrane using the four different sets.
The first $6$ states correspond to extrapolating the results for grids going from $N=10$ to $N=60$, with
$25$ unknown coefficients; the last two states correspond to extrapolating the results for grids going  
from $N=18$ to $N=60$, and with $21$ unknown coefficients. For a given state, the set with the asterisk 
corresponds to the minimal value taken by the least squares. The results which
do not converge to the exact value have been omitted.}
\end{table}

In  \cite{Berry87} Berry has devised an algorithm for obtaning successive approximations to 
the geometric properties $K_j$ of  a closed boundary $B$ given the lowest $N$ eigenvalues $E_n$. 
The partition function $\Phi(t) \equiv \sum_{n=1}^\infty e^{-E_n t}$ obeys an asymptotic expansion 
for small values of $t$
\beq
\Phi(t) \approx \frac{1}{t} \sum_{j=0}^\infty K_j t^{j/2}  \ ,
\eeq
where the coefficients $K_j$ are related to the geometric properties of $B$. For example
$K_0 = A/4 \pi$ and $K_1 = - \gamma L/8\sqrt{\pi}$. Using this asymptotic expansion Berry has 
obtained accelerated expressions for the geometrical constants of $B$. In particular for the area
of $B$ he has found the approximant (eq.(20) of \cite{Berry87})
\beq
A_m(t) = \frac{2\pi t}{m!} \sum_{n=1}^\infty e^{-\xi_n^2}  \ \xi_n^{m-1} \ H_{m+1}(\xi_n)
\label{Am}
\eeq
where $\xi_n \equiv \sqrt{E_n t}$. 

\begin{figure}[ht]
\begin{center}
\includegraphics[width=6cm]{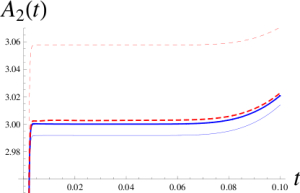}
\hspace{.5cm}
\includegraphics[width=6cm]{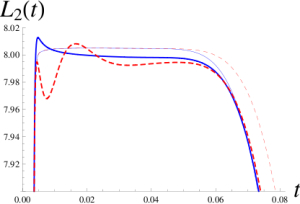}
\end{center}
\caption{Left Panel: The area approximant $A_2(t)$ obtained using the expression of Berry. The 
thin solid and dashed lines are obtained with the first $1000$ eigenvalues corresponding to the sets
$E_n^{(+)}$ and $E_n^{(-)}$ respectively. The bold solid and dashed lines correspond to the sets obtained
through an extrapolation from the original sets. Right Panel: The perimeter approximant $L_2(t)$ obtained with
the improved expression of Berry. The same sets of eigenvalues have been considered.}
\label{fig_berry}
\end{figure}

In the left panel of Fig.\ref{fig_berry} I show the area approximant $A_2(t)$, obtained using the expression
of Berry. The thin lines correspond to using the  the sets $E_n^{(+)}$ and $E_n^{(-)}$ (solid and dashed lines
respectively); the thick lines correspond to using the eigenvalues obtained from the extrapolation of  
the sets $E_n^{(+)}$ and $E_n^{(-)}$ (solid and dashed lines respectively). I call 
$\bar{E}_n^{(\pm)}$ the eigenvalues obtained extrapolating the eigenvalues $E_n^{(\pm)}$; the extrapolation
is carried out using the results obtained with grids with $N$ going from $N=48$ to $N=60$ and assuming
$E_n(N) \approx \bar{E}_n + \frac{\epsilon_n}{N}$. The approximants obtained with the extrapolated eigenvalues
provide excellent approximations to the area and perimeter of the membrane, as seen in Fig.\ref{fig_berry}.

\begin{figure}[ht]
\begin{center}
\includegraphics[width=7cm]{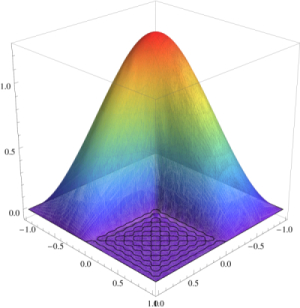}
\includegraphics[width=7cm]{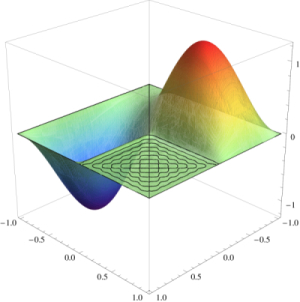}
\caption{First two eigenfunctions of the L-shaped membrane obtained using $N=30$. The black lines correspond
to the level $\psi(x,y) = 0$.}
\label{fig_lshaped3D0}
\end{center}
\end{figure}

Fig. \ref{fig_lshaped3D0} shows the first two eigenfunctions of the L shaped membrane obtained with a  grid 
corresponding to $N=30$. The solid lines appearing in the "forbidden region" correspond to the level $\psi(x,y) = 0$: 
the effect observed in the figure is due to the approximation of working with a finite number of grid points. In fact,
although a particular LSF vanishes on the points defining the grid, except on a particular point, where it reaches its 
maximum, it is non-zero elsewhere. This means that the numerical solution can take small values  even in the region
where the exact solution must vanish; however, the size of this effect decreases as the number of 
grid points is increased (taking into account that the computational load roughly increases as $N^4$). In the Appendix
we propose an alternative procedure which does not involve the diagonalization of larger matrices and which can be used to 
improved the results obtained with a given grid.

\section{The Africa drum}
\label{africa}

I will now examine the case of a membrane with an irregular shape. The application of the method proceeds exactly 
as in the case of the L-shaped membrane: once a grid is chosen, the points of the grid which are internal to the 
membrane are used to build a matrix representation of the Hamiltonian which, once diagonalized, provides the energies 
and wave functions of the problem.

\begin{figure}[t]
\begin{center}
\includegraphics[width=6cm]{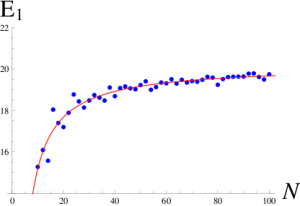}
\hspace{.5cm}
\includegraphics[width=6cm]{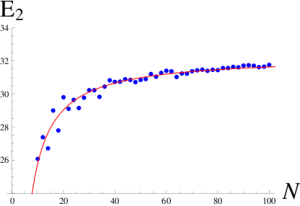}
\caption{Left: Energy of the fundamental mode of the Africa shaped membrane as a function of the number of grid points.
The continous line is the fit $E_1 = a + b/N$, with $a=20.1705$.
Right: Energy of the first excited mode of the Africa shaped membrane as a function of the number of grid points.
The continous line is the fit $E_1 = a + b/N$, with $a=32.2774$.}
\label{fig_africafit}
\end{center}
\end{figure}

As a paradigm of this class of membranes I have studied the vibrations of a drum with the shape of 
Africa. Unlike in the previous example the border does not cross the grid points, a feature which 
affects the precision of the results. The plots in Fig.\ref{fig_africafit} display the energies of the first two states
of the Africa drum for grids with different $N$ (the dots in the plots) and compare them with the best fit
obtained assuming that $E(N) = a + b/N$, where $a$ and $b$ are constants independent of $N$. The irregularity of the border
is reflected in the behavior of the eigenvalues which decay with $N$ but at the same time oscillate. 

\begin{figure}[t]
\begin{center}
\includegraphics[width=6cm]{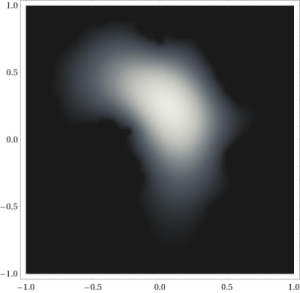}
\hspace{.5cm}
\includegraphics[width=6cm]{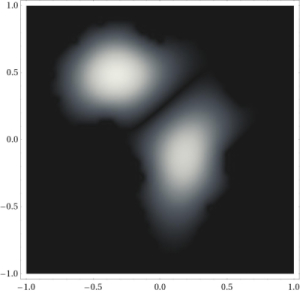}
\vspace{.5cm}
\includegraphics[width=6cm]{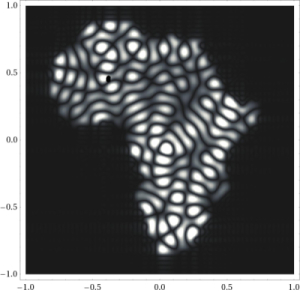}
\hspace{.5cm}
\includegraphics[width=6cm]{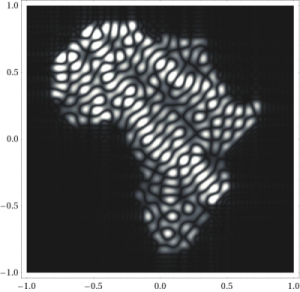}
\caption{Density plot for the fundamental state (upper left), first excited state (upper right),  $200^{th}$ excited state (lower left)
and  $300^{th}$ excited state (lower right) of the Africa shaped membrane.
In all plots the absolute value of the wave function is shown and a grid with $N=60$ is used.}
\label{fig_africa}
\end{center}
\end{figure}

\begin{figure}[t]
\begin{center}
\includegraphics[width=9cm]{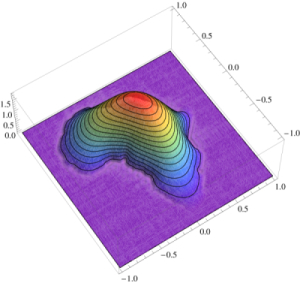}
\end{center}
\caption{Ground state of the Africa shaped membrane obtained using $N=60$.}
\label{fig_africa_1}
\end{figure}

In Fig.\ref{fig_africa} I show the density plot of four different states of the Africa drum, obtained using a grid
with $N=60$.  In Fig.\ref{fig_africa_1} I show the wave function of the ground state of the Africa drum, obtained
using a grid with $N=60$.

\section{Isospectral membranes}
\label{isospec}

In a classic paper dated 1966, \cite{Kac66}, Kac formulated an interesting question: whether it is possible to hear the
shape of a drum, meaning if the spectrum of frequencies of a given drum is unique to that drum or drums with different
shapes can have the same spectrum. The question was finally answered in 1992, when Gordon, Webb and Wolpert found a first
example of inequivalent drums having the same spectrum \cite{GWW92}. An experiment made by Sridhar and Kudrolli reported in \cite{SK94} 
used microwave cavities with the shape of the drums of \cite{GWW92} to verify the equality of the spectrum for the lowest 
$54$ states. More recently the same experiments have been carried out on isospectral cavities where the classical dynamics 
changes from pseudointegrable to chaotic \cite{DRSS03}.
Numerical calculations of the first few modes of the isospectral
drums found in \cite{GWW92} have been performed with different techniques: Wu, Sprung and Martorell~\cite{WSM95} have used a mode
matching method to calculate the first $25$ states of these drums and compared the results with those obtained with finite difference;
using a different approach Driscoll \cite{Dris97} has also calculated the first $25$ states obtaining results which are accurate 
to $12$ digits; Betcke and Trefethen \cite{BT05} have used their modified version of the method of particular solutions to obtain the
first three eigenvalues of these drums, reporting results which are slightly more precise than those of Driscoll.

I will now discuss the application of the present method to the calculation of the spectrum of these isospectral membranes: whereas in the
case of the L-shaped membrane the border of the membrane was sampled by the grid, regardless of the grid size (keeping $N$ even), in
the case of the isospectral membranes this happens only for grids where $N = 6 k$, with $k$ integer. It is important to restrict 
the calculation to this class of grids to avoid the oscillations observed in the case of the Africa membrane.
I have thus applied the method with grids ranging from $N=6$ to $N=120$\footnote{The numerical results presented in the case of the L-shaped 
membrane were obtained with a $40$-digit precision in the eigenvalues, using the command {\rm N[,40]} of Mathematica: in this case, 
since I need to resort to larger grids I have worked with less digits precision using the command $N[]$ in Mathematica.}. 

\begin{figure}[t]
\begin{center}
\includegraphics[width=7cm]{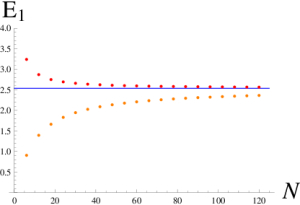}
\caption{Energy of the ground state of the first isospectral membrane as a function of the number of grid points $N$. 
The horizontal line is the precise result of \cite{BT05}. The set approaching the exact result from above (below) 
corresponds to $E_1^{(1+)}$ ($E_1^{(1-)}$).}
\label{fig_energyiso}
\end{center}
\end{figure}

The plot in Fig.\ref{fig_energyiso} displays the ground state energy of the first isospectral membrane calculated at different 
grid sizes. The horizontal line is the precise value given in \cite{BT05}.
The set approaching this value from above (below) corresponds to the application of the method rejecting (accepting) the grid points
falling on the border. The corresponding plot for the second isospectral membrane is almost identical and therefore it is not presented here.

In Table \ref{tableiso} I report the energies of the first 
$30$ states obtained using Richardson extrapolation \cite{BO} on the results for grids going from $N=66$ to $N=120$. The second and third columns
are the energy of the first isospectral membranes obtained with the sets which reject ($E_n^{(+)}$) or accept ($E_n^{(-)}$) the grid points 
falling on the border, which as seen in the case of the L-shape membrane provide a sequence of numerical values approaching the exact eigenvalue
from above and from below respectively. The last two columns report the analogous results for the second isospectral membrane.
Notice that some of the energies in  the third column are clearly incorrect.

\begin{figure}[t]
\begin{center}
\includegraphics[width=7cm]{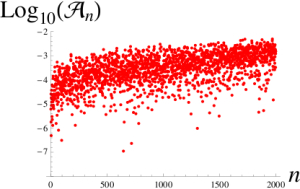}
\includegraphics[width=7cm]{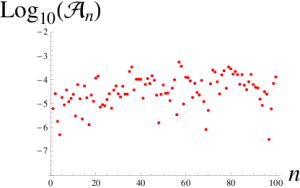}
\caption{Left panel: $\log_{10}$ of the asymmetry ${\cal A}_n \equiv (E_n^{(1+)} -E_n^{(2+)})/(E_n^{(1+)} +E_n^{(2+)})$
for the first $2000$ states of the isospectral membranes. $E_n^{(1+)}$ ($E_n^{(2+)}$ ) is the energy of the $n^{th}$ state of the 
first (second) membrane obtained using Richardson extrapolation of the grids with $N=114$ and $N=120$. Right panel: Blow-up of the 
previous plot for the first $100$ states.}
\label{fig_asymiso}
\end{center}
\end{figure}

A further empirical verification of the isospectrality of the two membranes is presented in Fig.\ref{fig_asymiso}, where I have
plotted the asymmetry ${\cal A}_n \equiv (E_n^{(1+)} -E_n^{(2+)})/(E_n^{(1+)} +E_n^{(2+)})$
for the first $2000$ states of the isospectral membranes. In this case $E_n^{(1+)}$ ($E_n^{(2+)}$ ) is the energy of the $n^{th}$ state of the 
first (second) membrane obtained using Richardson extrapolation of the grids with $N=114$ and $N=120$.

\begin{table}
\begin{center}
\begin{tabular}{|c|cc|cc|}
\hline
$n$ &  $E_n^{(1+)}$ & $E_n^{(1-)}$ &  $E_n^{(2+)}$ & $E_n^{(2-)}$ \\
\hline
1  &   2.537938184  &   2.537859157  &   2.537930157  &   2.537924672\\
2  &   3.655477379  &   3.655457482  &   3.655439267  &   3.655436933\\
3  &   5.175456364  &   5.175515223  &   5.1754891  &   5.175450085\\
4  &   6.53758046  &   6.537493542  &   6.537561774  &   6.537528448\\
5  &   7.247973684  &   7.248012453  &   7.247966062  &   7.248007219\\
6  &   9.209282216  &   9.209252596  &   9.20928929  &   9.209222008\\
7  &   10.59698943  &   10.59697476  &   10.59692509  &   10.59694683\\
8  &   11.54137149  &   11.54137651  &   11.54137016  &   11.54142735\\
9  &   12.33702671  &   12.33696554  &   12.33700655  &   12.33698898\\
10  &   13.05355072  &   13.0535318  &   13.05351736  &   13.05354013\\
11  &   14.31383084  &   14.31387457  &   14.31384362  &   14.31380888\\
12  &   15.87113023  &   15.87110476  &   15.87106608  &   15.8711794\\
13  &   16.94182893  &   -25414.06158  &   16.94177705  &   16.94177218\\
14  &   17.66507424  &   25448.66845  &   17.66503368  &   17.6650544\\
15  &   18.98079211  &   18.98079864  &   18.98083269  &   18.98081294\\
16  &   20.88240176  &   16.71191189  &   20.88233985  &   20.88246688\\
17  &   21.24773575  &   25.41816076  &   21.24772537  &   21.24764682\\
18  &   22.23265755  &   22.2326039  &   22.23265897  &   22.23262895\\
19  &   23.71129295  &   23.71135125  &   23.71127276  &   23.71130372\\
20  &   24.47925064  &   24.48080219  &   24.47920658  &   24.47934876\\
21  &   24.67406118  &   24.67245947  &   24.67401531  &   24.67403958\\
22  &   26.08011208  &   26.08090828  &   26.08008881  &   26.08012901\\
23  &   27.30391033  &   27.30298845  &   27.30390863  &   27.3039225\\
24  &   28.17508031  &   28.17506497  &   28.17506143  &   28.17505957\\
25  &   29.56976983  &   29.56970152  &   29.56975041  &   29.56905778\\
26  &   31.48308074  &   31.51241562  &   31.48304984  &   31.48393448\\
27  &   32.07624358  &   32.16454642  &   32.07622156  &   32.08008665\\
28  &   32.21611001  &   37.0118719  &   32.21605287  &   32.21393591\\
29  &   32.90535338  &   27.9888228  &   32.90537696  &   32.90354978\\
30  &   34.13633502  &   34.13929552  &   34.13632946  &   34.13632752\\
\hline
\end{tabular}
\label{tableiso}
\end{center}
\bigskip
\caption{First 30 eigenvalues of the isospectral membranes obtained with Richardson extrapolation of the results
obtained with grids from $N=66$ to $N=120$.}
\end{table}

\begin{figure}[t]
\begin{center}
\includegraphics[width=6cm]{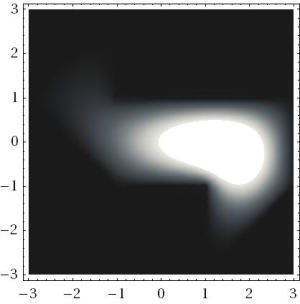}
\includegraphics[width=6cm]{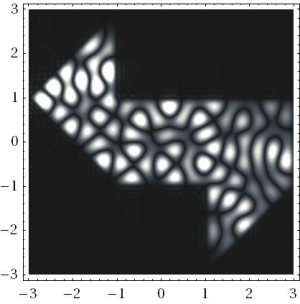}
\includegraphics[width=6cm]{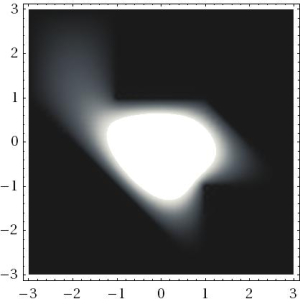}
\includegraphics[width=6cm]{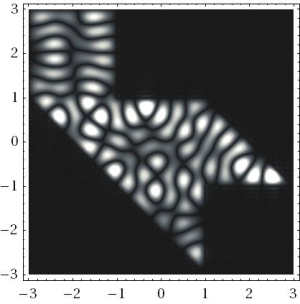}
\caption{Upper panel: Wave functions (absolute value) of the first isospectral membrane (ground state and $100^{th}$ excited state);
Lower panel: Wave functions (absolute value) of the second isospectral membrane (ground state and $100^{th}$ excited state). A grid with
$N=60$ is used.}
\label{isopsi}
\end{center}
\end{figure}

\section{An unusual drum}
\label{unusualdrum}

\begin{figure}[t]
\begin{center}
\includegraphics[width=7cm]{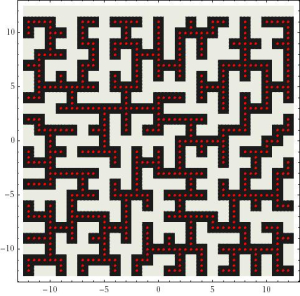}
\caption{The unusual drum considered by Trott \cite{Trott05}. The black area is the surface of the drum; the red points are the
collocation points corresponding to $N=50$.}
\label{fig_unusual}
\end{center}
\end{figure}

I will now consider a further example by looking at a particular membrane originally studied by Trott in \cite{Trott05}: this drum
is shown in Fig.\ref{fig_unusual} and consists of a total of $308$ units squares which are joined into a rather irregular 
form. Theoretical and experimental studies carried out on drums with fractal or irregular boundaries have shown that the wave excitations
for these drums are drastically altered \cite{SGM91,HSMV98,ERRPS99}: in particular, the Weyl law for these membranes is modified in a way 
which depends on the fractal dimension of the perimeter (see for example \cite{HYN95}), the so called Weyl-Berry-Lapidus conjecture.
Recently the vibrations of a uniform membrane contained in a Koch snowflake have been studied in two papers, \cite{NSS06,Banjai07}.

The paper by Trott is both interesting in its physical and mathematical content and as an example of the excellent capabilities of Mathematica
to handle heavy numerical calculations: as a matter of fact Trott uses a finite difference approximation of the Laplacian on a uniform grid 
and samples the membrane in $28521$ internal points. Explicit numerical values for the first $24$ modes are reported.

\begin{figure}[t]
\begin{center}
\includegraphics[width=7cm]{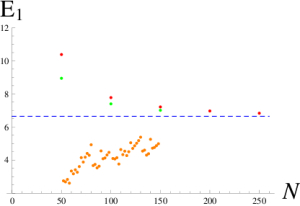}
\caption{Energy of the ground state of the unusual drum as a function of $N$. The horizontal line is the result of \cite{Trott05}; 
the points approaching the horizontal line from above correspond to configurations where the border is sampled by the collocation points
(and as discussed in the case of the L-shaped membrane are rejected). The green points correspond to the results obtained with the 
"mesh refinement" procedure described in the Appendix.}
\label{fig_unusual1}
\end{center}
\end{figure}

I have therefore considered the same problem using the LSF with  grids of different size (up to $N=250$ which leads to the same grid of
\cite{Trott05}). Figure \ref{fig_unusual1} displays the energy of the fundamental mode of this membrane as a function of the size of $N$.
The dashed horizontal line in the plot represents the result of  \cite{Trott05}, $E_1 = 6.64705$: the points on the upper part of the plot
correspond to $N$ going from $50$ to $250$, with intervals of $50$. For these particular values of $N$ the border of the membrane is sampled
by the grid and therefore more accurate results are expected. The grid points on the border are rejected, which leads to eigenvalues 
which approach the exact results from above, as seen in the previous examples. The points in the lower part of the plot correspond to grid sizes
varying from $N=52$ to $N=148$, excluding $N=100$: in this case the values approach the exact result from below, although in doing so 
they also oscillate reflecting the treatment of the border (a behaviour already observed in the case of the Africa membrane).
As mentioned above the finest grid corresponds to sampling the membrane on $28521$ internal points and therefore to working with
a $28521 \times 28521$ square matrix. Given that the matrix obtained with the LSF is a sparse symmetrix matrix, it is possible to deal 
efficiently with it in Mathematica, applying the Arnoldi method to extract a limited sequence of eigenvalues/eigenvectors.
The reader will notice that in this example I have not considered the set corresponding to accepting the grid points falling on the border, as
done in the case of the L-shaped and of the isospectral membranes: although this set provides a sequence of values which uniformly approach the
value at the continuum, the number of grid points sampled is quite large  because of the large perimeter of
the membrane. For example, for $N=100$, this set samples the membrane on $7029$ points, compared with the $N= 3801$ points used in the 
other set.

The Figure also displays the improved ground state energies obtained using the "mesh refinement" procedure described in the Appendix (the
three green points): the eigenvector for a given grid is extrapolated to a finer grid rejecting contributions in the "forbidden region" 
(i.e. falling outside the border of the membrane). The improved energy estimate corresponds to the expectation value of the Hamiltonian 
in this state and thus it requires no diagonalization. The results displayed in the figure correspond to extrapolation to a grid which
is twice finer.

\section{Bound states in the continuum}
\label{BIC}

It is well known that the spectrum of the Laplacian with Dirichlet boundary conditions may contain bound states even for 
open geometries, in correspondence of crossings or bendings of the domain. For example, Schult et al.\cite{Schult89} have studied
the problem of two crossed wires, of infinite length, showing that such geometry supports exactly one bound state, localized at the 
crossing. Avishai and collaborators have also proved the existence of a bound state in the broken strip configuration for arbitrarily 
small angles, see \cite{Avis91} (more recently Levin has proved the existence of one bound state in the broken strip for any 
angle of the strip \cite{Levin04}). Goldstone and Jaffe \cite{GoldJaff92} have given a variational proof of the existence of a bound state for an
infinite tube in two and three dimensions, provided that the tube is not straight. Other interesting configurations which support
bound states in the continuum have been studied by Trefethen and Betcke \cite{TB06}.

The example which I will consider here is somehow related to the crossed wires configuration studied by Schult at al. I
have considered a set of horizontal and vertical wires, of neglegible trasverse dimension, which are contained in a square 
box of size $2$. Calling $\bar{n}$ the number of wires in each dimension, $\bar{n}^2$ is the number of crossings between these wires 
(for simplicity the wires are assumed to be equally spaced). This configuration can be easily studied in the present 
collocation approach, by sampling the wires on a grid and by then diagonalizing the Hamiltonian obtained following this procedure.
The resulting energies calculated in this way will clearly depend on the spacing of the collocation grid, $h$, and 
diverge as $h$ is sent to zero. To obtain finite results one needs to multiply these eigenvalues by $h^2$, which eliminates the divergence
caused by the shrinking of the transverse dimension. Following this procedure I have studied different configurations, corresponding
to choosing different value of $\bar{n}$ (going from $\bar{n}=1$ to $\bar{n}=4$) and  I have found that a given configuration 
has precisely the same number of bound states as the number of crossing. These bound states happen to be almost exactly degenerate and
correspond to wave functions which are localized on the vertices.

In Table \ref{table_cross} I report the energy (multiplied by $h^2$) of the bound states and of the first unbound state ($E_{gap}$) 
for the different configurations. These results have been obtained using a fine grid corresponding to $h=1/300$ and show that the bound
states are precisely $\bar{n}^2$ as anticipated and they are essentially degenerate; the energy of the bound states and of the gap are 
also found to be almost insensitive to $\bar{n}$, which can be interpreted as a sign of confinement of a state to the crossings. I have also
checked the dependence of these results upon $N$ (or equivalently upon $h$) observing that the energies can be fitted excellently as 
$E = a + b/N^2$; for example in the case of the ground state of the configuration with $\bar{n}=4$ I have obtained: $E=2.59874- 44.6364/N^2$.

\begin{table}
\begin{center}
\begin{tabular}{|c|cccc|}
\hline
$\bar{n}$ &  $1$ & $2$ &  $3$ & $4$ \\
\hline
$h^2 E_1$  & 2.59873 &   2.59871  &  2.59867  &   2.59862 \\
$h^2 E_2$  &   -  &   2.59873     &  2.59869   &  2.59864\\
$h^2 E_3$  &   -  &   2.59873  &   2.59869  &     2.59864\\
$h^2 E_4$  &   -  &   2.59876  &   2.59872  &     2.59867\\
$h^2 E_5$  &   -  &   -  &   2.59873  &   2.59868\\
$h^2 E_6$  &   -  &   -  &   2.59873  &   2.59868\\
$h^2 E_7$  &   -  &   -  &   2.59876  &   2.59871\\
$h^2 E_8$  &   -  &   -  &   2.59876  &   2.59871\\
$h^2 E_9$  &   -  &   -  &   2.59880  &   2.59874\\
$h^2 E_{10}$  &   -  &   -  &   -  &      2.59874\\
$h^2 E_{11}$  &   -  &   -  &   -  &      2.59875\\
$h^2 E_{12}$  &   -  &   -  &   -  &      2.59877\\
$h^2 E_{13}$  &   -  &   -  &   -  &      2.59877\\
$h^2 E_{14}$  &   -  &   -  &   -  &      2.59881\\
$h^2 E_{15}$  &   -  &   -  &   -  &      2.59881 \\
$h^2 E_{16}$  &   -  &   -  &   -  &      2.59887\\
\hline
$h^2 E_{gap}$ & 3.28997  & 3.29006        &     3.29019         &  3.29035\\
\hline
\end{tabular}
\end{center}
\bigskip
\caption{Energies of the bound states for configurations with different number of crossings, 
using $N=600$, corresponding to a spacing $h = 1/300$.}
\label{table_cross}
\end{table}

\begin{figure}[t]
\begin{center}
\includegraphics[width=9cm]{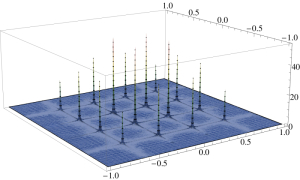}
\caption{Wave function of the ground state of the configuration for $\bar{n}=4$ using $N=500$.}
\label{fig_grate3D}
\end{center}
\end{figure}

In Fig.\ref{fig_grate3D} I have plotted the wave function of the ground state of the configuration corresponding to $\bar{n}=4$ using
a grid with $N=500$. The wave function is clearly localized at the crossings between the wires. Similar behaviour is observed
for the remaining $15$ bound states.

\section{Collocation with conformal mapping}
\label{conformal}

The examples considered in the previous Sections show that it is possible to obtain the spectrum of the negative Laplacian over 
regions of arbitrary shape by using a collocation scheme, where the boundary conditions need not to be explicitly enforced on the border.
Clearly, the precision of this approach should improve if the boundary conditions would be enforced exactly on the border of the
membrane. One way of achieving this result is by mapping conformally the shape of the membrane into a square (or a rectangle), on whose border
the LSF obey Dirichlet boundary conditions. I will discuss explicitly two examples of how this is done.

\subsection{Circular membrane}

As a first example I consider a circular homogeneous membrane, which is exactly solvable (see for example \cite{FW80}) 
and therefore it can be a useful tool to test the precision of the present method.

\begin{figure}[t]
\begin{center}
\includegraphics[width=5cm]{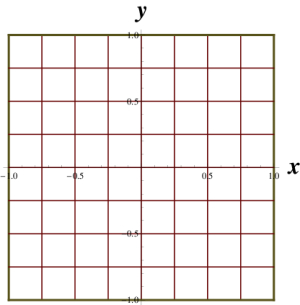}
\hspace{.5cm}
\includegraphics[width=5cm]{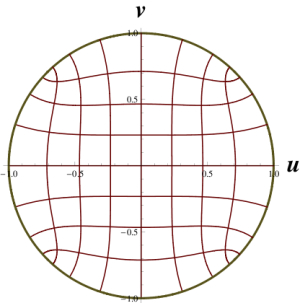}
\end{center}
\caption{Unit square in the $z$ plane and the corresponding unit cirle in the $w$ plane reached through the trasformation (\ref{eq_sch}).} 
\label{fig_circle}
\end{figure}

The function
\beq
f(z) = e^{-\frac{3 i \pi }{4}} \ {\rm sn}\left(\left.z \ F\left(\left.\sin^{-1}\left(e^{-\frac{i \pi }{4}}\right)\right|-1\right)\right|-1\right)
\label{eq_sch}
\eeq
maps the unit square in the $w$ complex plane into the unit circle in the complex $z$ plane, as seen in Fig.~\ref{fig_circle}.
Under this mapping the original equation, 
\beq
-\Delta  \psi(w) = \lambda \psi(w)
\label{circle1}
\eeq
with Dirichlet boundary conditions on the unit circle, is mapped to 
\beq
-\Delta  \chi(z) = \lambda \sigma(z) \chi(z)
\label{circle2}
\eeq
with Dirichlet boundary conditions on the unit square.  Here $\sigma(z) \equiv \left|\frac{dw}{dz}\right|^2$
and eq.~(\ref{circle2}) describes the vibrations of a non-uniform square membrane. Although in the previous Sections 
I have restricted the application of the method to the case of uniform membranes of arbitrary shapes, the method can be 
applied also to inhomogenous membranes straightforwardly. Let me briefly mention how this is done.
As a first step eq.~(\ref{circle2}) may be written in the equivalent form 
\beq
-\frac{1}{\sigma(z)} \Delta  \chi(z) = \lambda \chi(z) \ .
\label{circle3}
\eeq

The operator $\hat{O} \equiv \frac{1}{\sigma(z)} \Delta$ is evaluated on a uniform grid in the $z$-plane using the 
Little Sinc Functions (LSF). The action of the operator over a product of sinc functions can be calculated 
very easily, as explained in the previous Sections. To make the discussion simpler, I restrict to the equivalent 
one dimensional operator and make it act over a single LSF:
\beq
-\frac{1}{\sigma(x)} \frac{d^2}{dx^2} s_k(h,N,x) &=& - \sum_{jl} \frac{1}{\sigma(x_j)} c^{(2)}_{kl} s_j(h,N,x) s_l(h,N,x) \nonumber \\
&\approx& - \sum_{j} \frac{1}{\sigma(x_j)} c^{(2)}_{kj} s_j(h,N,x) \ .
\eeq

The matrix representation of the operator over the grid may now be read explicitly from the expression above. The reader should notice
that the matrix will not be symmetric unless the density is constant~\footnote{In general the calculation of the eigenvalues and eigenvectors
of non--symmetric matrices is computationally more demanding than for symmetric matrices of equal dimension. }.

Using this approach I have considered grids with $N= 10,20,\dots,80$ and I have have calculated the first four even-even eigenvalues, 
which are shown in Table \ref{tablecircle}. Taking into account the symmetry of problem I have used symmetrized LSF, 
which obey mixed boundary conditions (Dirichlet at one end and Neumann at the other hand): in this way, for a given value of the $N$ 
a grid of $(N/2)^2$ points is used. As mentioned before the exact eigenvalues for this problem are known (the zeroes of the Bessel functions):
these are reported in the last row.

In fig.~\ref{fig_energy_circle} I have plotted the lowest eigenvalue of the circular membrane corresponding to different $N$ and I have fitted 
these points using functions like $c_0+c_1/N^r$, with $r=3,4,5$ (the dashed, solid and dotted lines
in the plot). This plot shows that the leading (non--constant) behaviour of the numerical energy for $N \gg 1$ is
$1/N^4$.

Taking into account this behaviour I have considered the quantity
\beq
\Xi_Q \equiv \sum_{k=1}^8 \left[ \alpha_1 - \sum_{n=2}^Q \frac{\alpha_n}{(10 k)^{n+2}} \right]^2  \ ,
\eeq
where $Q=8$ and I have obtained the coefficients $\alpha_n$ by minimizing $\Xi_Q$ (notice that this expression
takes into account the leading $1/N^4$ behaviour just discussed).
The row marked as $LSQ_{8}$ displays the quite precise results obtained following this procedure.

\begin{figure}[t]
\begin{center}
\includegraphics[width=7cm]{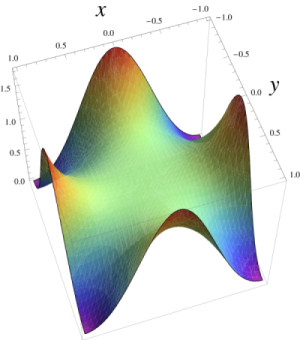}
\end{center}
\caption{Density of the inhomogeneous square membrane isospectral to the homogeneous circular membrane.}
\label{fig_density_circle}
\end{figure}

\begin{figure}[t]
\begin{center}
\includegraphics[width=7cm]{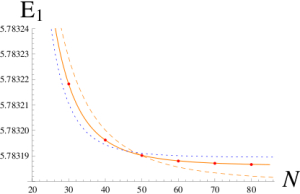}
\end{center}
\caption{Energy of the ground state of the circular membrane. The dashed, solid and dotted lines 
correspond to fits using functions like $c_0+c_1/N^r$, with $r=3,4,5$ respectively.}
\label{fig_energy_circle}
\end{figure}

\begin{table}
\begin{center}
\begin{tabular}{|c|cccc|}
\hline
$N$ & $E_1$ & $E_2$ & $E_3$ & $E_4$ \\
\hline
10     & 5.785633618  &  26.46056162 & 30.55061880 & 57.88187288 \\
20     & 5.783347847  &  26.37986506 & 30.47598468 & 57.60026669 \\
30     & 5.783218252  &  26.37564237 & 30.47217988 & 57.58626207 \\
40     & 5.783196213  &  26.37493961 & 30.47155075 & 57.58397911 \\
50     & 5.783190167  &  26.37474851 & 30.47138009 & 57.58336363 \\
60     & 5.783187992  &  26.37468004 & 30.47131902 & 57.58314408 \\
70     & 5.783187059  &  26.37465074 & 30.47129291 & 57.58305035 \\
80     & 5.783186606  &  26.37463653 & 30.47128023 & 57.58300497 \\
\hline
LSQ$_8$ & 5.783185971  & 26.37461646 & 30.47126209 & 57.58294087 \\
\hline
Exact  & 5.783185962  &  26.37461642 & 30.47126234 & 57.58294090 \\
\hline
\end{tabular}
\end{center}
\bigskip
\caption{Even-even spectrum of the circular membrane: first four eigenvalues}
\label{tablecircle}
\end{table}

I would like to discuss briefly a different issue. In \cite{Gott88} Gottlieb has used the Moebius transformation
\beq
f_g(z) = (z-a)/(1-a z)
\label{moebius}
\eeq
to map the unit circle onto itself. This mapping transforms the homogenoeous
Helmoltz equation for a circular membrane into the inhomogeneous Helmoltz 
equation for a circular membrane with density
\beq
\rho(x,y) = \left| f_g'(z)\right|^2 = \rho_0  \frac{(1-a)^2}{\left[(1-a x)^2+a^2 y^2\right]^2} \ .
\eeq

\begin{figure}[ht]
\begin{center}
\includegraphics[width=5cm]{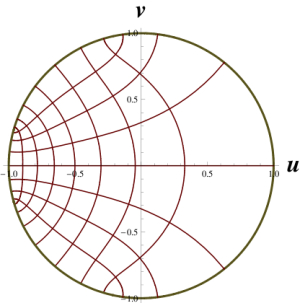}
\hspace{.5cm}
\includegraphics[width=5cm]{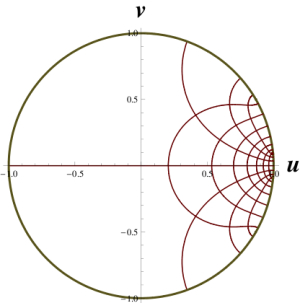}
\end{center}
\caption{Grid obtained with the Moebius map corresponding to $a=0.5$ (left) and $a=-0.8$ (right). } 
\label{fig_circle2}
\end{figure}

\begin{figure}[ht]
\begin{center}
\includegraphics[width=7cm]{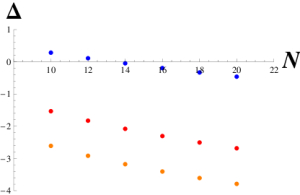}
\end{center}
\caption{$\Delta \equiv Log_{10} (E_N-E_{exact})$ using three values of $a$ ($a=0$, $0.4$ and $0.8$ from bottom to top). } 
\label{fig_circle3}
\end{figure}

Gottlieb uses this result to conclude that membranes corresponding to different densities, i.e. different values of $a$, 
are isospectral, thus providing  a negative answer to the famous question ``Can one hear the shape of a drum?'', 
posed by Kac in \cite{Kac66}. I wish to move our discussion on computational grounds: for a given $a$ 
the mapping of eq.~(\ref{moebius}) deforms the grid inside the unit circle; as $a$ is changed, the grid points move,
as shown in Fig.~\ref{fig_circle2}. The case $a=0$ is plotted in the right panel of Fig.~\ref{fig_circle}.
Clearly, if the density of the membrane is constant, or symmetric with respect to the center, one expects that $a=0$
provide the best grid. In Fig.~\ref{fig_circle3} I have plotted the logarithm of the difference between the 
approximate and exact energy for the ground state of a circular membrane, $\Delta \equiv Log_{10} (E_N-E_{exact})$, using 
three values of $a$ ($a=0$, $0.4$ and $0.8$). These numerical results confirm the prediction made: stated in different terms
one can conclude that {\sl for a given problem one can improve the numerical accuracy of a calculation by selecting an 
optimal grid among those obtained through a conformal map of the region onto itself.} 
The optimization of the parameter $a$ depending on the specific problem considered is in the same spirit of the variational 
approach used in \cite{Amore06,Amore07a,Amore07b} and could provide a useful computational tool to boost the precision of the
results.

\subsection{Circular waveguide}

The second example of application of conformal mapping to the solution of the Helmholtz equation is taken 
from the paper of Kuttler and Sigillito \cite{KS84} (this problem was also studied earlier by Moler, in  ref.[101] of  
\cite{KS84}).

\begin{figure}[ht]
\begin{center}
\includegraphics[width=5cm]{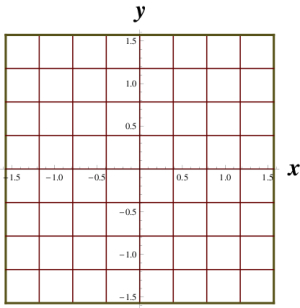}
\hspace{.5cm}
\includegraphics[width=5cm]{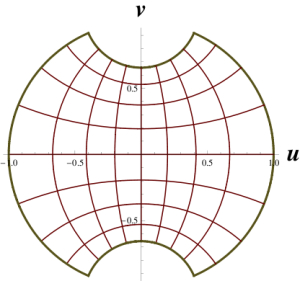}
\end{center}
\caption{Square in the $z$ plane and corresponding region in the $w$ plane, reached through the conformal
map $w = \tan \frac{z}{2}$.}
\label{fig_kutt}
\end{figure}

In Fig.\ref{fig_kutt} two regions of the plane are displayed: the left plot corresponds to a square of side $\pi$ 
centered on the origin in the $z = x+i y$ plane; the right plot corresponds to a circular waveguide with 
circular ridges in the $w = u + i v$ plane.  The function $w = \tan \frac{z}{2}$ maps the first region 
into the second one. 

As I have shown for the case of the circular membrane, the homogeneous Helmoholtz equation over the second region may be
transformed into an inhomogeneous Helmholtz equation over the square:
\beq
-\Delta  U(z) = \lambda \sigma(z) U(z)  \ .
\label{kutt2}
\eeq
In the present case  $\sigma(z) \equiv \left|\frac{dw}{dz}\right|^2 = (\cos x+\cosh y)^2$ and Dirichlet 
boundary conditions are assumed on the borders of the two regions.  

In Tables 1,2 and 3 of their paper, Kuttler and Sigillito report different estimates for the first $12$
even-even eigenvalues, obtained using different approaches. In Table 2 they also apply Richardson extrapolation
to the eigenvalues obtained with finite difference. In the case of the ground state of this membrane they
also mention the precise value obtained by Moler using the method of point matching
\beq
\lambda_1 = 7.5695769
\eeq

\begin{table}
\begin{center}
\begin{tabular}{|c|cccc|}
\hline
$N$ &  $E_1$ & $E_2$ & $E_3$ & $E_4$ \\
\hline
10  &   7.575738906  &   29.35369905  &   44.93667650  &   68.99532514\\
20  &   7.569970385  &   29.12882337  &   44.84592568  &   67.91298030\\
30  &   7.569654735  &   29.11799633  &   44.84124500  &   67.86357065\\
40  &   7.569601533  &   29.11623444  &   44.84047707  &   67.85592485\\
50  &   7.569586991  &   29.11575957  &   44.84026961  &   67.85390710\\
60  &   7.569581767  &   29.11559019  &   44.84019553  &   67.85319500\\
70  &   7.569579528  &   29.11551787  &   44.84016389  &   67.85289283\\
80  &   7.569578441  &   29.11548286  &   44.84014857  &   67.85274711\\
\hline
{\rm LSQ$_8$} & 7.569576902 & 29.11543343 & 44.84012692 & 67.85254236 \\
{\rm LSQ$_7$} & 7.569576902 & 29.11543343 & 44.84012692 & 67.85254236 \\
\hline
\hline
$N$ &  $E_5$ & $E_6$ & $E_7$ & $E_8$ \\
\hline
10  &   76.36327173  &   105.8649443  &   127.5818229  &   147.6128111\\
20  &   74.57343676  &   104.7105731  &   123.4501146  &   137.5136748\\
30  &   74.51254455  &   104.6448241  &   123.2916952  &   137.1508752\\
40  &   74.50340797  &   104.6345417  &   123.2690972  &   137.1033030\\
50  &   74.50101871  &   104.6318226  &   123.2633192  &   137.0914797\\
60  &   74.50017885  &   104.6308625  &   123.2613110  &   137.0874237\\
70  &   74.49982321  &   104.6304550  &   123.2604661  &   137.0857295\\
80  &   74.49965192  &   104.6302584  &   123.2600608  &   137.0849203\\
\hline
{\rm LSQ$_8$} & 74.49941161 & 104.6299823 & 123.2594952 & 137.0837970 \\
{\rm LSQ$_7$} & 74.49941160 & 104.6299823 & 123.2594952 & 137.0837959 \\
\hline
\hline
$N$ &  $E_9$ & $E_{10}$ & $E_{11}$ & $E_{12}$ \\
\hline
10  &   152.6380731  &   175.0500571  &   202.7827432  &   229.6150278\\
20  &   147.1852075  &   177.5293898  &   193.4167694  &   213.4362048\\
30  &   147.1167888  &   177.2332164  &   193.0075863  &   212.8440230\\
40  &   147.1064916  &   177.1901085  &   192.9541314  &   212.7718374\\
50  &   147.1038082  &   177.1790645  &   192.9409198  &   212.7546507\\
60  &   147.1028673  &   177.1752263  &   192.9364009  &   212.7488774\\
70  &   147.1024696  &   177.1736121  &   192.9345164  &   212.7464938\\
80  &   147.1022783  &   177.1728379  &   192.9336173  &   212.7453635\\
\hline
{\rm LSQ$_8$} & 147.1020103 & 177.1717582 & 192.9323707 & 212.7438068 \\
{\rm LSQ$_7$} & 147.1020110 & 177.1717573 & 192.9323707 & 212.7438134 \\
\hline
\end{tabular}
\end{center}
\bigskip
\caption{Even-even eigenvalues of the problem of eq.~(\ref{kutt2}) using collocation with Little Sinc Functions (LSF).}
\label{table2}
\end{table}

\begin{figure}[ht]
\begin{center}
\includegraphics[width=7cm]{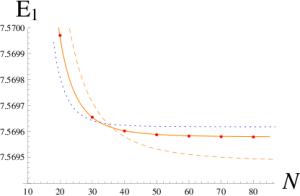}
\end{center}
\caption{Energy of the ground state of the circular waveguide. The dashed, solid and dotted lines 
correspond to fits using functions like $c_0+c_1/N^r$, with $r=3,4,5$ respectively.}
\label{fig_kutt2}
\end{figure}

In Table \ref{table2} I report the even-even eigenvalues of the eq.~(\ref{kutt2}) obtained using collocation
with different values of $N$. The results corresponding to the ground state are plotted in Fig.
\ref{fig_kutt2} and fitted using functions like $c_0+c_1/N^r$, with $r=3,4,5$ (the dashed, solid and dotted lines
in the plot). This plot proves that the leading (non--constant) behaviour of the numerical energy for $N \gg 1$ is $1/N^4$,
as for the circular membrane.

The results in the Table have also been extrapolated using a least square approach
\beq
\Xi_Q \equiv \sum_{k=1}^8 \left[ \alpha_1 - \sum_{n=2}^Q \frac{\alpha_n}{(10 k)^{n+2}} \right]^2  \ ,
\eeq
where $Q=7,8$ and $\alpha_n$ are coefficients which are obtained by minimizing $\Xi_Q$. Notice that this expression
takes into account the leading $1/N^4$ behaviour just discussed.
The rows marked as $LSQ_{7,8}$ display the results obtained following this procedure (the comparison between the results for
$Q=7$ and $Q=8$ gives an indication over the precision reached): in particular the energy of the ground state 
reproduces all the digits of the result obtained by Moler. It is also remarkable that the energies obtained
with the conformal-collocation method decrease monotonically when the number of collocation points is increased (the only
exception is represented by the $E_{10}$ for $N=10$, probably due to the limited number of collocation points).

As a technical remark, one should notice that the results corresponding to a given value of $N$ are obtained using a set 
of $N/2$ symmetric (even) functions for each direction, thus reducing the computation load by a factor of $4$. 
The results displayed in this table should be compared with the analogous results of Table 2 of \cite{KS84}, 
which were obtained using finite difference.

\begin{figure}[t]
\begin{center}
\includegraphics[width=6cm]{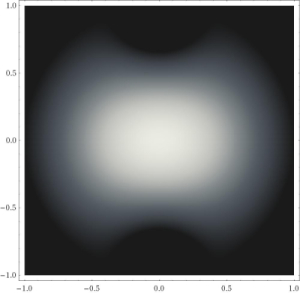}
\includegraphics[width=6cm]{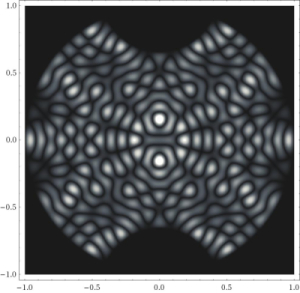}
\includegraphics[width=6cm]{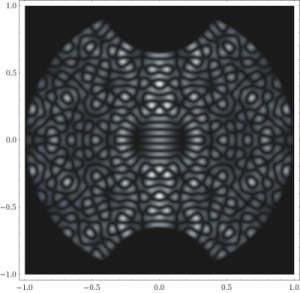}
\includegraphics[width=6cm]{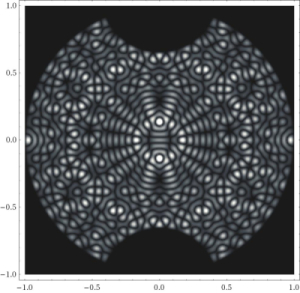}
\caption{Upper panel: Even-even wave functions (absolute value): ground state and $100^{th}$ excited state of the circular waveguide;
Lower panel: Even-even wave functions (absolute value): $200^{th}$ and $300^{th}$ excited states of the circular waveguide. A grid with
$N=80$ is used.}
\label{fig_kuttpsi1}
\end{center}
\end{figure}

\section{Conclusions}
\label{conclusion}

In this paper I have used a collocation method based on LSF to obtain the numerical solutions of the Helmholtz equation over two-dimensional
regions of arbitrary shape. A large number of examples has been studied, illustrating the great potentialities of the present method.
Among the principal virtues of this method I would like to mention its generality (it can be applied to membranes of arbitrary shapes, 
including inhomogeneous membranes, and to the Schr\"odinger equation -- although I have not done this in the present paper), 
its simplicity (the matrix representation of the Helmholtz operator is obtained directly by collocation, and therefore it does not 
require the calculation of integrals) and the possibility of combining it with a conformal mapping, as done in the last Section.
In this last case, a rapid convergence to the exact eigenvalues is observed as the number of grid points is increased. In the case where the
border is not treated exactly it has also been observed that the method provides monotonous sequences of approximations to the exact eigenvalue
either from above or from below.
Readers interested to looking at more examples of application of this method may find useful to check the gallery of images which
can be found at 
\begin{verbatim*}
http://fejer.ucol.mx/paolo/drum
\end{verbatim*}

\appendix

\section{Mesh refinement}

Although the collocation method described in this paper allows one to obtain precise solutions to the Helmholtz equation over domains 
of arbitrary shape, in general the Dirichlet boundary conditions are not enforced exactly over all the boundary. 
As discussed in Section \ref{conformal} the best approach consists of introducing a conformal map, which allows one to go from  
the original problem to an inhomogenous Helmoltz problem over a square: in such case the Dirichlet boundary conditions are imposed exactly
and rapid convergence to the exact solutions is observed.
In general, however, finding such a conformal map can be a difficult task and therefore the first approach may be more appealing. 
I will discuss here a simple procedure to ``refine'' the results obtained by direct collocation of the Helmholtz equation over the
grid. The fundamental observation is that the LSF that we have used do vanish 
on the grid points on the border and external to the membrane, but they are nonzero in all the other points external to the membrane.
Therefore the cumulative effects of the LSF internal to the membrane can be seen also outside the membrane, although it will
tend to disappear as the number of grid points is increased. This solution, to increase the number of grid points, may be the most 
obvious but it is certainly not appealing computationally, since increasing the number of grid points strongly increases the 
computational cost (remember that the number of matrix elements grows as $N^4$). However we can use much simpler procedure,
which does not require any additional diagonalization. Call $N$ the parameter defining the size of the grid: a point in this grid 
is described by the direct product of the LSF in the $x$ and $y$ directions. In the Dirac notation we write 
\beq
\langle x, y|k,k' \rangle_h \approx s_k(h,x) s_{k'}(h,y) \ ,
\eeq
assuming for simplicity that the grid has the same spacing in both directions. Let us now concentrate on one of the LSF, 
say the one in the $x$ direction: we take a finer grid, with a spacing $h' =  h/l$, where $l$ is a integer. 
The new grid contains now $(l N-1)$ points, including obviously the original grid points. However, it is clear that 
the original LSF can be decomposed in the new grid as
\beq
s_k(h,x) = \sum_{j=-l N/2+1}^{l N/2-1} s_k(h,\bar{x}_j) \ s_j(h/l,x)  ,
\eeq
where $\bar{x}_j = 2 L j/(l N)$ are the new grid points. Notice that this relation is exact.

The wave function of the $n^{th}$ state obtained from the
diagonalization of the $(N-1) \times (N-1)$ hamiltonian reads 
\beq
&& \psi_n(x,y) = \frac{1}{h} \ \sum_{K} v_K^{(n)} \ s_{k(K)}(h,x) \ s_{k'(K)}(h,y)  \nonumber \\
&=& \frac{1}{h} \ \sum_{K} v_K^{(n)} \ \sum_{j=-l N/2+1}^{l N/2-1} s_{k(K)}(h,\bar{x}_j) \ s_j(h/l,x)  \
\sum_{j'=-l N/2+1}^{l N/2-1} s_{k'(K)}(h,\bar{y}_{j'}) \ s_{j'}(h/l,y) \ ,  \nonumber
\eeq
where $v^{(n)}$ is the $n^{th}$ eigenvector. Clearly $\psi_n(x,y)$ differs from $0$ even in points of the refined
grid which fall outside the membrane profile. We introduce a new matrix whose elements
are given by
\beq
\eta_{jj'} = \left\{\begin{array}{c}
0 \ \  if \ (\bar{x}_j,\bar{y}_{j'}) \ \notin \ {\cal B} \\
1 \ \ if \ (\bar{x}_j,\bar{y}_{j'}) \ \in \ {\cal B} \\
\end{array}\right.
\eeq
and rewrite the wave function ``purged'' on the refined grid as
\beq
\bar{\psi}_n(x,y) &=& \frac{{\cal N}}{h} \ \sum_{j=-l N/2+1}^{l N/2-1} \sum_{j'=-l N/2+1}^{l N/2-1} \tilde{V}_{jj'} 
\ s_j(h/l,x)   \ s_{j'}(h/l,y)  \nonumber
\eeq
where
\beq
\tilde{V}_{jj'} \equiv  \eta_{jj'} \sum_{K} v_K^{(n)} \  s_{k(K)}(h,\bar{x}_j) \ s_{k'(K)}(h,\bar{y}_{j'}) 
\eeq
and ${\cal N}$ is a normalization constant that ensures that 
\beq
\int_{{\cal B}} \psi^2_n(x,y) dx \ dy = 1 \ .
\eeq

It is easy to show that 
\beq
{\cal N} = \frac{l}{\sqrt{ \sum_{jj'} \tilde{V}_{jj'}^2 }} \ .
\eeq

To simplify the notation I define:
\beq
{\cal V}_{jj'} \equiv \frac{\cal N}{l} \tilde{V}_{jj'}
\eeq
and thus write:
\beq
\bar{\psi}_n(x,y) &=& \frac{l}{h} \ \sum_{j=-l N/2+1}^{l N/2-1} \sum_{j'=-l N/2+1}^{l N/2-1} {\cal V}_{jj'} 
\ s_j(h/l,x)   \ s_{j'}(h/l,y)  \nonumber
\eeq

On the other hand we may also calculate the expectation value of the Hamiltonian in this state
\beq
&& \langle \hat{H} \rangle_n = -\int_{\cal B} \bar{\psi}_n(x,y) \ \Delta \bar{\psi}_n(x,y) \ dx \ dy  \nonumber \\
&=&  - \sum_{jj'rr'ss'} \frac{{\cal V}_{jj'} {\cal V}_{rr'}}{h^2} \left[ \bar{c}^{(2)}_{rs} \delta_{r's'} + \bar{c}^{(2)}_{r's'} \delta_{rs} \right]  
\ \int_{\cal B}   s_j(h/l,x)   \ s_{j'}(h/l,y) s_s(h/l,x)   \ s_{s'}(h/l,y) \nonumber \\
&=&  - \sum_{jj'rr'} {\cal V}_{jj'} {\cal V}_{rr'}  
\left[ \bar{c}^{(2)}_{rj} \delta_{r'j'} + \bar{c}^{(2)}_{r'j'} \delta_{rj} \right]    \nonumber \\
&=&  -  \sum_{jj'r} \bar{c}^{(2)}_{rj} \left[ {\cal V}_{jj'} {\cal V}_{rj'} +  {\cal V}_{j'j} {\cal V}_{j'r} \right] 
\eeq
where $\bar{c}^{(2)}$ is the matrix for the second derivative on the refined grid. An example of application of this procedure is
shown in Fig.\ref{fig_unusual1}.

\verb''\section*{References}
\verb''

\end{document}